\documentclass[twocolumn,amsmath,amssymb,longbibliography,superscriptaddress,aps]{revtex4-2}
\usepackage{graphicx,bm}
\usepackage{appendix}
\usepackage{color}
\usepackage{mathtools}
\usepackage{soul}
\usepackage{float}
\usepackage{braket}

\begin{document}

\title{Exceptional point in a trimer chain of oscillators with a quadratic driving}
\author{M~Shoufie Ukhtary} 
\affiliation{Research Center for Quantum Physics, National Research and Innovation Agency (BRIN), South Tangerang 15314, Indonesia}
\email{m.shoufie.ukhtary@brin.go.id}
\author{Albert Andersen} 
\affiliation{Research Center for Quantum Physics, National Research and Innovation Agency (BRIN), South Tangerang 15314, Indonesia}
\affiliation{Department of Physics, Faculty of Mathematics and Natural Sciences,
Universitas Indonesia, Depok 16424, Indonesia}
\author{Donny Dwiputra}
\author{M. Jauhar Kholili}
\affiliation{Research Center for Quantum Physics, National Research and Innovation Agency (BRIN), South Tangerang 15314, Indonesia}


\begin{abstract}
Exceptional points of a dissipative chain of three coupled oscillators (trimer), which is driven by quadratic photon, are investigated.  The exceptional points emerge from the coalescence of both eigenvalues and eigenvectors of the dynamical matrix that describes the first moments of the trimer. At the exceptional point, we found that the optical spectrum is split into two peaks, instead of a conventional single peak, as in the case of a single oscillator. In particular, the positions of these peaks correspond to the natural frequency of the trimer in a \textit{closed system}, which depends only on the coupling strength. Furthermore, after passing the exceptional point, the peak positions do not change, which can be used to estimate the coupling strength between oscillators. 
\end{abstract}
\date{\today}
\maketitle
\section{Introduction}
\label{sec:int}
Exceptional point (EP) has been the key feature of a non-Hermitian system, in which gain and loss play an important role in the realization of EP \cite{wiersig2020review,miri2019exceptional,ashida2020non,ozdemir2019parity}. At EP, both the eigenvalues and the eigenvectors of an operator describing the dynamics of a system coalesce into a singularity followed by a transition in dynamics \cite{wiersig2020review,miri2019exceptional,ashida2020non,ozdemir2019parity,heiss1998collectivity, downing2021exceptional}. For example, the population dynamics in a quantum system changes from oscillatory in time to exponentially increasing after passing EP \cite{downing2021exceptional, downing2023quantum}. The collapse of both eigenvalues and eigenvectors presents a strong response of the system to external perturbation \cite{wiersig2020review}. In particular, this strong response to perturbation has been applied for developments of sensitive sensor working in the vicinity of EP \cite{wiersig2020review, hodaei2017enhanced,chen2017exceptional,wiersig2014enhancing,wiersig2016sensors,kuo2020collectively}. 

The dynamics of an open system is often described by the Lindbladian master equation \cite{manzano2020short,breuer2002theory,gardiner2015quantum}. The nonunitary effects are incorporated into the master equation by the so-called dissipators associated with their jump operators. From the master equation, the associated dynamical matrix of statistical moments can be derived. This matrix fully describes the evolution of the system over time, since the eigenvalues give the frequency of the system. Moreover, the coalesce of the eigenvalues and their corresponding eigenvectors of this matrix, thus, corresponds to EP of the system \cite{downing2021exceptional,downing2023quantum,arkhipov2020liouvillian,perina2022quantum}. To investigate the presence of EP experimentally, the optical spectrum or the power spectrum is measured \cite{kavokin2017microcavities,downing2023parametrically,downing2023resonance,breuer2002theory,gardiner2015quantum}. Passing EP usually changes the number of peaks and the lineshape of the spectrum. Normally, the peak-to-peak separation corresponds to the splitting of the eigenvalues. Therefore, approaching EP, the separation becomes smaller since at least two eigenvalues are merging. In a simple system consisting of two objects, the spectrum changes from doublet to singlet as we pass EP \cite{downing2023parametrically}. 

Typically, EP is investigated in an interacting multipartite system, where the coupling constant, gain, or loss rate determine EP. Downing et al. considered a case of a single oscillator driven parametrically, from which they found EP determined by the driving strength and the loss rate \cite{downing2023parametrically}. Furthermore, they found that the optical spectrum changes from doublet to singlet when passing EP \cite{downing2023parametrically}. In this work, we investigate EP in a multipartite system consisting of three interacting quantum oscillators driven parametrically, as shown in Fig. \ref{fig:tr}. Our aim is to study the effect of coupling on EP and the corresponding optical spectrum. At EP, we found that the optical spectrum is split into two peaks, instead of conventional single peak as in the case of a single oscillator. In particular, the positions of these peaks correspond to the natural frequency of the trimer in a \textit{closed system}, which depends only on the coupling strength. Therefore, we can apply our system to estimate the coupling strength using the optical spectrum at EP.

\section{Model}
\label{sec:met}

\begin{figure}[t]
\begin{center}
\includegraphics[width=75mm]{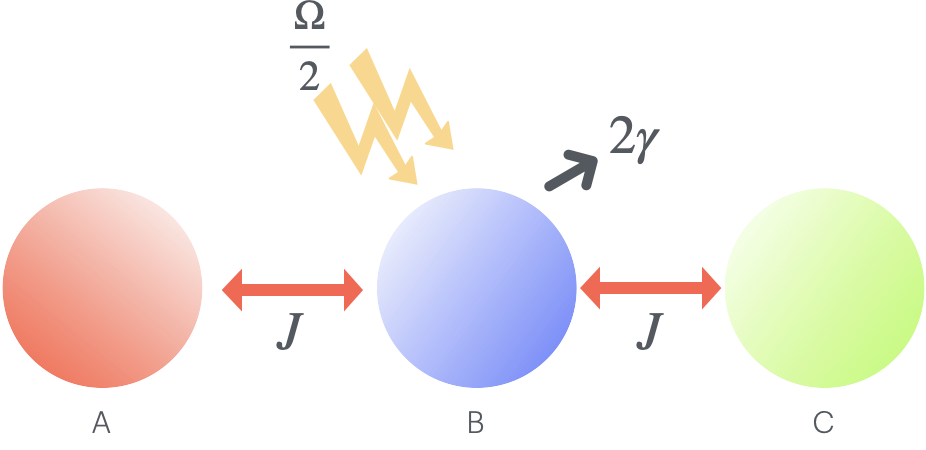}
\caption{ A chain of interacting trimer. The system is driven quadratically with strength of $\Omega/2$ on oscillator B.  We allow dissipation with rate of $2\gamma$ on oscillator B. Oscillator B is coupled with two other oscillators with coupling constant of $J$.}\label{fig:tr}
\end{center}
\end{figure}
The trimer consists of three interacting oscillators as illustrated in Fig. \ref{fig:tr}, with level spacing of each oscillator $\omega_i\, (i=a,b,c)$. We consider a quadratic driving on oscillator B, which is coupled with two other oscillators with a coupling constant of $J$. In this case, we consider the system that obtains energy from external quadratic driving, but the energy dissipates to the environment incoherently. The system Hamiltonian is expressed as follows:
\begin{align}
    H&=\omega_a a^\dag a + \omega_b b^\dag b + \omega_c c^\dag c + J(a^\dag b + a b^\dag+ b^\dag c +bc^\dag)\nonumber\\
    &+\frac{\Omega}{2}\left(b^\dag b^\dag e^{-i2\omega_Dt}+ bbe^{i2\omega_Dt}\right).\label{eq:hm}
\end{align}
To remove the explicit time dependence of the Hamiltonian, we move to a frame rotating with the driving by using the unitary operator of $U=e^{-i\omega_Dt(a^\dag a+b^\dag b+c^\dag c)}$. The transformed Hamiltonian is given as follows:
\begin{align}
    H&= \Delta_a a^\dag a + \Delta_b b^\dag b + \Delta_c c^\dag c + J(a^\dag b + a b^\dag+ b^\dag c +bc^\dag)\nonumber\\
    &+\frac{\Omega}{2}\left(b^\dag b^\dag + bb\right),\label{eq:hmt}
\end{align}
where $\Delta_i=\omega_i-\omega_D$ is the detuning frequency of oscillator-$i$. For simplicity, we assume that oscillators A and C are resonant with driving, while we allow detuning of the frequency of oscillator B as $\Delta\equiv\Delta_b$.

The system is coupled to the environment through dissipation in oscillator B with a dissipation rate of $2\gamma$, as illustrated in Fig. \ref{fig:tr}. This allows us to compare it with the case of a dissipative oscillator driven parametrically studied by Downing et al. and to understand the impact of the coupling. The dynamics of the system is governed by the following master equation for density matrix $\rho$,
\begin{align}
    \partial_t\rho&=-i[H,\rho]+\gamma\left(2b\rho b^\dag-b^\dag b\rho-\rho b^\dag b \right),\label{eq:me}
\end{align}
where the first term describes the Hermitian evolution of the system, while the second term gives the dissipation of the system. Using Eq. \eqref{eq:me} and the trace property $\partial_t\langle O\rangle=\textrm{Tr}[O\partial_t\rho]$, we derive the following differential equation for the first moment of the system,
\begin{align}
    i\partial_t\Psi=
\mathcal{H}\Psi,\label{eq:sh}
\end{align}
where $\Psi=(\langle a\rangle, \langle b\rangle, \langle c\rangle ,\langle a^\dag\rangle,\langle b^\dag\rangle , \langle c^\dag\rangle)^T$
and \begin{align} 
\mathcal{H}=\begin{pmatrix}0&J&0&0&0&0\\J &\Delta-i\gamma& J&0&\Omega&0\\
0& J&0&0&0&0\\
0&0&0&0&-J&0\\
0&-\Omega&0&-J&-\Delta-i\gamma&-J\\
0&0&0&0&-J&0 \end{pmatrix}\label{eq:H}
\end{align}
The dynamics of the system is determined by the matrix $\mathcal{H}$, which can be understood as the effective Hamiltonian of the system. The eigenvalues of $\mathcal{H}$, denoted by $\lambda$,  give the frequencies for the evolution of the system. Due to the presence of $\gamma$, the eigenvalues are generally complex, which highlights the relaxation of the system to a steady state. We found analytical solutions for the eigenvalues as follows: 
\begin{align}
\lambda_{1,2}&=0,\quad\lambda_{3,4}=\frac{1}{2} \left(-i (\gamma +\Pi)\pm\omega_- \right)\nonumber\\
\lambda_{5,6}&=\frac{1}{2} \left(-i (\gamma -\Pi)\pm\omega_+ \right),\label{eq:ev}
\end{align}
where we define,
\begin{align}
\Pi\equiv \sqrt{\Omega^2 -\Delta^2  },\,\omega_\pm&\equiv\sqrt{Z\pm2 \gamma  \Pi },\, Z&\equiv 8 J^2-\gamma ^2-\Pi^2.
\end{align}
From Eq. \eqref{eq:ev}, we found that there are three distinct driving strengths $\Omega$ that generate exceptional points (EPs), which are given by the zeros of the square roots argument. The EPs are given by the following driving strength,
\begin{align}  
\Omega^{(1)}_\textrm{EP}&=\Delta,\quad\Omega^{\pm}_\textrm{EP}=\sqrt{\Delta ^2+(\gamma\pm2\sqrt{2}J)^2}\label{eq:ep}
\end{align}
We only consider the case of steady state, which is represented by negative imaginary part of the eigenvalue. The positive imaginary part of the eigenvalue leads to spectral collapse in the case of a closed oscillator driven quadratically \cite{downing2023parametrically,downing2023quantum}. The critical driving, above which the steady state is not present, is given by $\Omega_c\equiv \sqrt{\gamma^2+\Delta^2}$. Using this expression, we can limit our parameters to $2J^2\le \gamma^2$, so that the steady state is present. From Eq. \eqref{eq:ep}, we obtain an EP at the driving strength independent of $J$, $\Omega^{(1)}_\textrm{EP}$. Interestingly, this driving strength also gives EP in the case of a single driven oscillator \cite{downing2023parametrically}. Therefore, at any coupling strength, we can always find the EP. In the subsequent sections, we will investigate this EP $\Omega^{(1)}_\textrm{EP}$, which will give us indication whether the driven oscillator is coupled to other oscillators or isolated. 

To investigate the properties of the corresponding EP, we derive the first-order correlation function $g_b^{(1)}(\tau)$, from which we can further calculate the optical spectrum of the emitted photon from the driven oscillator B. The normalized correlation function is defined as follows \cite{kavokin2017microcavities,downing2023parametrically,breuer2002theory,gardiner2015quantum}:
\begin{align}
    g_b^{(1)}(\tau)=\lim_ {t\rightarrow\infty}\frac{\langle b^\dag (t)b(t+\tau)\rangle}{\langle b^\dag (t)b(t)\rangle},
\end{align}
where $\lim_ {t\rightarrow\infty}\langle b^\dag (t)b(t)\rangle\equiv \langle b^\dag b^\dag\rangle_\textrm{ss} $ is the steady-state population of oscillator B. The steady-state population is solved from the differential equation for second moments, similar to Eq. \eqref{eq:sh}. The resulted steady-state populations are expressed as follows:
\begin{align}
    \langle a^\dag a\rangle_\textrm{ss}&=\langle c^\dag c\rangle_\textrm{ss}=\frac{\Omega ^2}{4 \left(\gamma ^2+\Delta ^2-\Omega ^2\right)}\\
    \langle b^\dag b\rangle_\textrm{ss}&=\frac{\Omega ^2}{2 \left(\gamma ^2+\Delta ^2-\Omega ^2\right)}\\
    \langle b^\dag b^\dag\rangle_\textrm{ss}&=\frac{i (\gamma  \Omega +i \Delta  \Omega )}{2 \left(\gamma ^2+\Delta ^2-\Omega ^2\right)}.
\end{align}
$g_b^{(1)}(\tau)$ is derived using the quantum regression theorem \cite{khan2022quantum,breuer2002theory,gardiner2015quantum}. $g_b^{(1)}(\tau)$ is obtained by solving the following differential equation,
\begin{align}
\partial_\tau v=-i\mathcal{H}v\label{eq:qrt},
\end{align}
where $\mathcal{H}$ is given by Eq. \eqref{eq:H} and $v=(\langle b^\dag(t) a(t+\tau)\rangle, \langle b^\dag(t)b(t+\tau)\rangle, \langle b^\dag(t)c(t+\tau)\rangle ,\langle b^\dag(t)a^\dag(t+\tau)\rangle,\langle b^\dag(t)b^\dag(t+\tau)\rangle , \langle b^\dag(t)c^\dag(t+\tau)\rangle)^T$. In solving Eq. \eqref{eq:qrt}, the initial conditions $(\tau=0)$ are given by the steady-state solutions of the corresponding second moments. Therefore, the first-order correlation function is expressed as follows:
\begin{widetext}
\begin{align}
g_b^{(1)}(\tau)=&\frac{i}{ 2\Pi  \Omega^3  }\Bigg[\frac{e^{-\frac{1}{2} (\Gamma_+  \tau)}}{\omega_-} \left(\Omega ^3+i \alpha_+  \beta_+ \Omega \right) \left(\beta_- \omega_- \cos \left(\frac{\omega_-}{2}\tau\right)-\left(\alpha_+ \beta_--i \Omega ^2\right) \sin \left(\frac{\omega_-}{2}\tau\right)\right)\nonumber\\
&-\frac{e^{-\frac{1}{2} (\Gamma_-  \tau)}}{\omega_+} \left(\Omega ^3+i \alpha_+ \beta_- \Omega \right) \left(\beta_+ \omega_+\cos \left(\frac{ \omega_+}{2}\tau\right)+\left(-\alpha_+ \beta_+ +i \Omega ^2\right) \sin \left(\frac{\omega_+}{2}\tau\right)\right)\Bigg],
\end{align}
where $\Gamma_{\pm}\equiv \gamma \pm \Pi,\,\alpha_\pm\equiv\gamma \pm i \Delta,\,\beta_\pm\equiv \Delta \pm i \Pi $.
The optical spectrum is calculated by taking the Fourier transform of $g_b^{(1)}(\tau)$. Since $g_b^{(1)}(\tau)$ consists of four independent terms, the Fourier transform can be taken separately for each term. The total spectrum $S(\omega)$ is expressed as follows \cite{kavokin2017microcavities,downing2023parametrically}:
\begin{align}
    S(\omega)\equiv&\frac{1}{\pi}\textrm{Re}\int\limits_{0}^\infty g_b^{(1)}(\tau)e^{i \omega \tau} d\tau=\frac{1}{\pi}\textrm{Re}f_s(\omega),
\end{align}
where
\begin{align}
    f_s(\omega)=&\frac{i}{2\Pi\Omega^3}\left[\frac{\Omega ^3+i \alpha_+  \beta_+ \Omega}{\omega_-}\left[\beta_-\omega_-F_1-(\alpha_+\beta_--i\Omega^2)F_2\right]-\frac{\Omega ^3+i \alpha_+  \beta_- \Omega}{\omega_+} \left[\beta_+\omega_+F_3-(\alpha_+\beta_+-i\Omega^2)F_4\right]\right]\label{eq:s},
\end{align}
and
\begin{align}
F_1=&\frac{\frac{\Gamma_+}{2}-i\omega}{\left(\frac{\omega_-}{2}\right)^2+\left(\frac{\Gamma_+}{2}-i\omega\right)^2},\quad F_2=\frac{\frac{\omega_-}{2}}{\left(\frac{\omega_-}{2}\right)^2+\left(\frac{\Gamma_+}{2}-i\omega\right)^2}\nonumber\\
F_3=&\frac{\frac{\Gamma_-}{2}-i\omega}{\left(\frac{\omega_+}{2}\right)^2+\left(\frac{\Gamma_-}{2}-i\omega\right)^2},\quad F_4=\frac{\frac{\omega_+}{2}}{\left(\frac{\omega_+}{2}\right)^2+\left(\frac{\Gamma_-}{2}-i\omega\right)^2}\nonumber.
\end{align}
\end{widetext}
As we can see from Eq. \eqref{eq:s}, the optical spectrum can be separated into four Lorentzian lineshapes. The spectrum is expressed further as follows:
\begin{figure*}[t]
\begin{center}
\includegraphics[width=180mm]{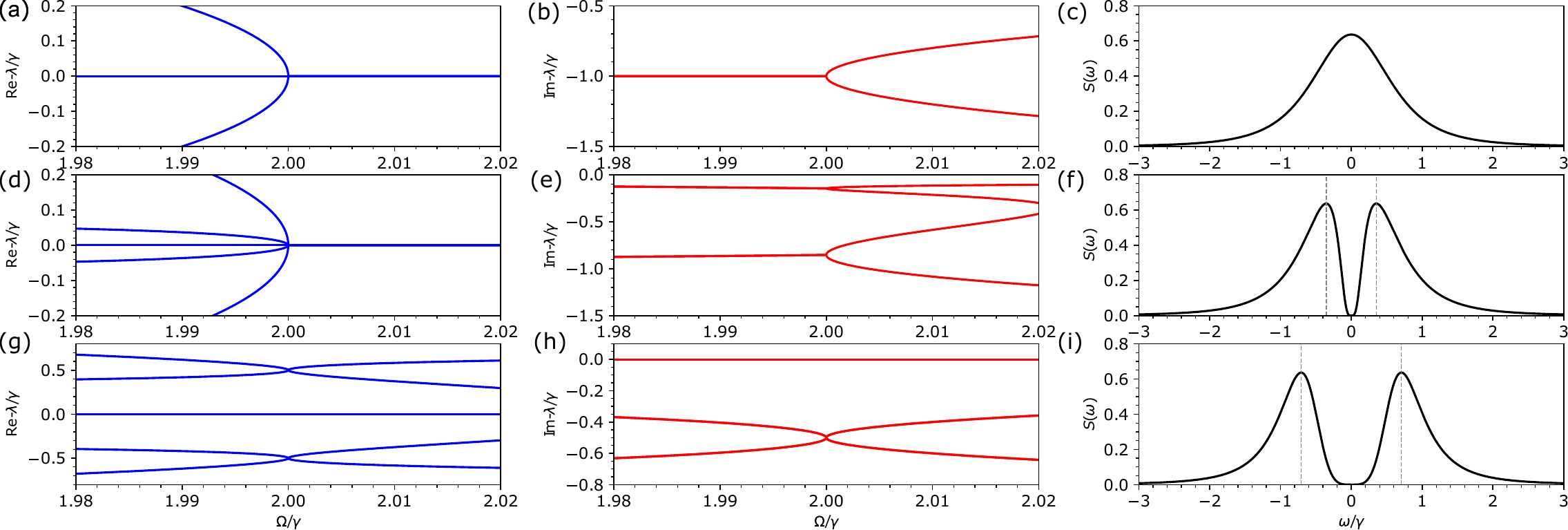}
\caption{(a) and (b) The real and imaginary parts of $\lambda$ in the vicinity of the EP , $\Omega_\textrm{EP}^{(1)}$, respectively. (c) The optical spectrum at the EP $\Omega^{(1)}_\textrm{EP}$, $\Omega=\Delta$. Here we use $\Delta=2\gamma$ and $J=0$. (d) - (f) The same as (a) - (c) but for $J=0.25\gamma$ and (g) - (i) for $J=0.5\gamma$. The dashed lines in the optical spectra correspond to the peak positions $\pm\sqrt{2}J$. In all cases, we set $\gamma=1$}\label{fig:s}
\end{center}
\end{figure*}
\begin{widetext}
\begin{align}
    S(\omega)=\frac{2 \gamma  \omega ^4 \left(\gamma ^2-\Pi ^2\right)}{\pi  \left(\omega ^4 \left(\gamma ^4+2 \gamma ^2 \left(\omega ^2-\Pi ^2\right)+\left(\Pi ^2+\omega ^2\right)^2\right)+16 J^8-32 J^6 \omega ^2+8 J^4 \omega ^2 \left(\gamma ^2+\Pi ^2+3 \omega ^2\right)-8 J^2 \omega ^4 \left(\gamma ^2+\Pi ^2+\omega ^2\right)\right)}
\end{align}
\end{widetext}
\section{Results and Discussion}
Before looking at the optical spectrum, let us investigate $\lambda$ at the exceptional point with $\Omega=\Omega_\textrm{EP}^{(1)}$. At an exceptional point, two or more eigenvalues and the corresponding eigenvectors coalesce.  From Eq. \eqref{eq:ev}, we have two distinct $\lambda$ when $\Omega=\Delta$. The $\lambda_3$ and $\lambda_5$ coalesce with eigenvalues of,
\begin{align}
\lambda^{(1)}_\textrm{EP,3}=\lambda^{(1)}_\textrm{EP,5}=-i\frac{\gamma}{2} +\sqrt{2J^2-(\frac{\gamma}{2})^2}
\end{align}
and the corresponding coalesced eigenvectors of,
\begin{align}
\mathbf{u}_3=\mathbf{u}_5=\left(1,\frac{\lambda^{(1)}_\textrm{EP,3}}{ J},1,1,-\frac{\lambda^{(1)}_\textrm{EP,3} }{ J},1\right)^T.
\end{align}
On the other hand, the $\lambda_4$ and $\lambda_6$ coalesce with eigenvalues of,
\begin{align}
\lambda^{(1)}_\textrm{EP,4}=\lambda^{(1)}_\textrm{EP,6}=-i\frac{\gamma}{2} -\sqrt{2J^2-(\frac{\gamma}{2})^2}
\end{align}
and the corresponding coalesced eigenvectors of,
\begin{align}
\mathbf{u}_4=\mathbf{u}_6=\left(1,\frac{\lambda^{(1)}_\textrm{EP,4}}{ J},1,1,-\frac{\lambda^{(1)}_\textrm{EP,4} }{ J},1\right)^T.
\end{align}
Therefore, there are two distinct EPs when we set $\Omega=\Delta$ and $J>0$, as shown by Fig. \ref{fig:s}. On the other hand, in the case of $J=0$, all eigenvalues and eigenvector coalesce into one EP creating a higher order EP as shown in Figs. \ref{fig:s} (a) and (b).

In the case of $2J^2<(\gamma/2)^2$, $\lambda$ is purely imaginary as shown by Figs. \ref{fig:s} (d) and (e). The two EPs are clearly shown in the imaginary part of $\lambda$, while the real part of $\lambda$ of all branches vanishes. In the case of $2J^2>(\gamma/2)^2$, $\lambda$ is complex, with the two EPs clearly shown in the real part of $\lambda$ as shown in \ref{fig:s} (g) and (h). In all cases, $J$ determines the separation of the two EPs in $\lambda$, with separation of $\Delta\lambda_\textrm{EP}=2\sqrt{2J^2-(\gamma/2)^2}$.
\begin{figure*}[t]
\begin{center}
\includegraphics[width=180mm]{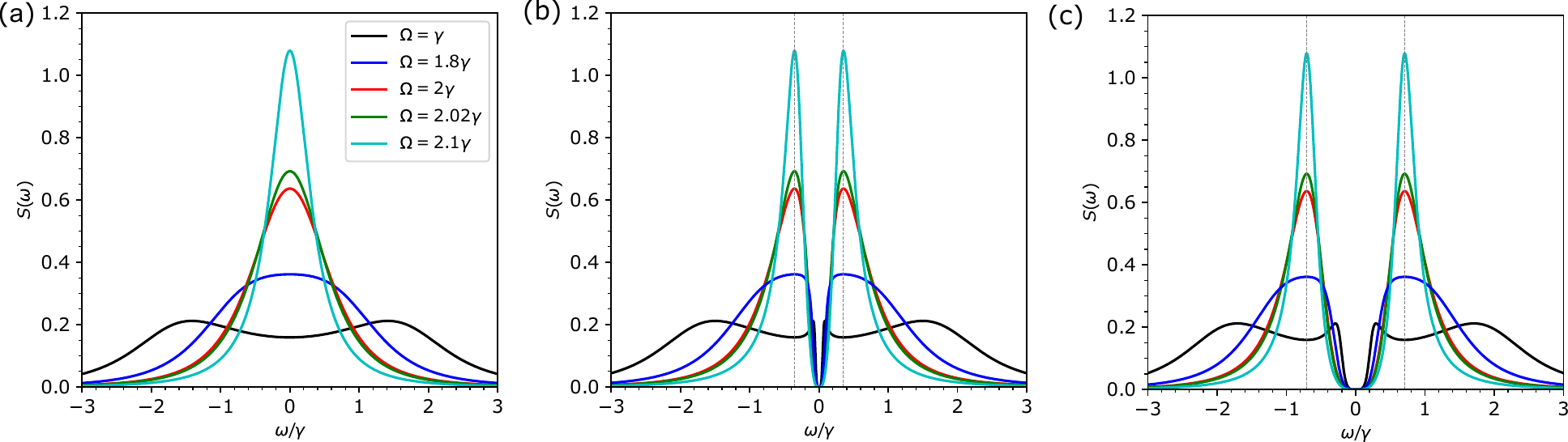}
\caption{Optical spectra for several $\Omega$. (a) For $J=0$. (b) For $J=0.25\gamma$. (c) For $J=0.5\gamma$. In all cases, we set $\Delta=2\gamma$ and $\gamma=1$. The dashed lines in the optical spectra correspond to the peak positions $\pm\sqrt{2}J$. The EP is given by $\Omega=\Delta=2\gamma$.}\label{fig:s2}
\end{center}
\end{figure*}
Let us investigate the optical spectrum corresponding to these EPs. 
As we mentioned before, these EPs are generated by the same driving strength as in the case of single oscillator $\Omega^{(1)}_\textrm{EP}=\Delta$, which does not depend on $\gamma$ and $J$. Therefore, it is intriguing to probe the optical spectrum at these EPs and look into how the coupling affects the optical spectrum.

At exactly $\Omega=\Delta$, the optical spectrum is expressed simply as
\begin{widetext}
\begin{align}
    S_\textrm{EP,1}(\omega)=&
    \frac{2 \gamma ^3 \omega ^4}{\pi  \left(\left(\left(\frac{\gamma }{2}\right)^2+(\omega -\omega^{(1)}_\textrm{EP,1})^2\right) \left(\left(\frac{\gamma }{2}\right)^2+(\omega+\omega^{(1)}_\textrm{EP,1})^2\right)\right)^2}
    =\frac{2 \gamma ^3 \omega ^4}{\pi  \left(\gamma ^2 \omega ^2+\left(\omega ^2-2 J^2\right)^2\right)^2}\label{eq:s1}
\end{align}
\end{widetext}
for the case of $2J^2>(\gamma/2)^2$, where we define $\omega^{(1)}_\textrm{EP,1}\equiv \sqrt{2J^2-(\gamma/2)^2}$. The peaks of the spectrum are located at $\omega_{p}=\pm\sqrt{(\gamma/2) ^2+(\omega^{(1)}_\textrm{EP,1})^2}=\pm\sqrt{2}J$. For the case of $2J^2<(\gamma/2)^2$, where we define $\omega^{(2)}_\textrm{EP,1}\equiv \sqrt{(\gamma/2)^2-2J^2}$, the spectrum is expressed as follows:
\begin{widetext}
\begin{align}
    S_\textrm{EP,1}(\omega)=&
    \frac{2 \gamma ^3 \omega ^4}{\pi  \left(\left((\gamma -2 \omega^{(2)}_\textrm{EP,1})^2+4 \omega ^2\right) \left((\gamma +2 \omega^{(2)}_\textrm{EP,1})^2+4 \omega ^2\right)\right)^2}=\frac{2 \gamma ^3 \omega ^4}{\pi  \left(\gamma ^2 \omega ^2+\left(\omega ^2-2 J^2\right)^2\right)^2}\label{eq:s2}
\end{align}
\end{widetext}
The peaks are located at $\omega_{p}=\pm\sqrt{(\gamma/2) ^2-(\omega^{(2)}_\textrm{EP,1})^2}=\pm\sqrt{2}J$. Interestingly, in both cases, the final expressions of the spectrum are the same as given by Eqs. \eqref{eq:s1} and \eqref{eq:s2}, giving the same peak positions, which do not depend on $\gamma$. 

At the zero coupling $J=0$, we recover the spectrum for a driven single oscillator \cite{downing2023parametrically},
\begin{align}
S_\textrm{EP}^\textrm{single}=\frac{2 \gamma ^3}{\pi  \left(\gamma ^2+\omega ^2\right)^2},
\end{align}
where the peak is located at $\omega=0$ as shown in Fig. \ref{fig:s} (c). This peak position simply corresponds to $\lambda=-i\gamma$ as shown in Figs. \ref{fig:s} (a) and (b), where all branches coalesces at zero real part of $\lambda$. The purely imaginary $\lambda$ gives the broadening of the spectrum centered on $\omega=0$. On the other hand, when the coupling is finite, the spectrum is split into two symmetric peaks, with a peak separation of $\Delta\omega_\textrm{EP}=2\sqrt{2}J$. This peak separation is clearly shown in Figs. \ref{fig:s} (f) and (i), where we use $J=0.25\gamma$ and $J=0.5\gamma$, respectively. From Eqs. \eqref{eq:s1} and \eqref{eq:s2}, we also find that the intensity of the spectrum vanishes for $\omega=0$ as soon as $J$ is finite, which distinguishes it from the spectrum of a single oscillator. More interestingly, this peak separation is different from the separation of EP eigenvalues $\Delta\lambda_\textrm{EP}$. In particular, in the case of $2J^2<(\gamma/2)^2$, where the separation of EP appears in the imaginary $\lambda$, the two peaks emerge, instead of one peak like the case of the single oscillator in (c). 

To better understand the optical spectrum, we calculate the spectrum for several $\Omega$ as shown in Fig. \ref{fig:s2}. The spectrum for a single oscillator is shown in Fig. \ref{fig:s2}(a). There appear two symmetric peaks when $\Omega=1/2\Omega^{(1)}_\textrm{EP}=\gamma$, which correspond to the two branches of the real part of $\lambda$. Approaching EP, the two peaks merge into a single broad peak. A clear single Lorentzian spectrum starts to appear at the EP and stays centered at $\omega=0$ for $\Omega>\Omega^{(1)}_\textrm{EP}=\Delta$, as expected since the real part of $\lambda$ vanishes for all branches. Unlike a single oscillator, when $J$ is finite, we have four peaks instead of two, as shown in Figs. \ref{fig:s2} (b) and (c), which corresponds to the four distinct real parts of $\lambda$ when $\Omega=1/2\Omega^{(1)}_\textrm{EP}=\gamma$. Approaching the EP, each of the two peaks merges into a single peak at $\omega=\pm\sqrt{2}J$ and stays at the same positions even for $\Omega>\Omega^{(1)}_\textrm{EP}=\Delta$.  Thus, the change from quadruplet to doublet in the optical spectrum is the signature of passing EPs in this coupled oscillator.  In addition to the change of the number of peaks as a signature of passing EPs, the appearance of two peaks at EP implies the presence of a coupling in the oscillator (compared with the case of a single oscillator).

This peculiar position of the peaks comes from the presence of a dispersive term $\omega^4$ on the numerator in Eqs. \eqref{eq:s1} and \eqref{eq:s2}. Without $\omega^4$, the expression of the optical spectrum will be pure Lorentzian, with the peak position corresponding to the real part of $\lambda$. In particular, we would have a single peak at $\omega=0$ for $2J^2<(\gamma/2)^2$ instead of a doublet in Fig. \ref{fig:s}(f). In the presence of $\omega^4$, the spectrum at $\omega=0$ vanishes, which shifts the peak positions of the pure Lorenztian spectrum to $\omega=\pm\sqrt{2}J$. Interestingly, these peak position corresponds to the natural frequency of the trimer system in a closed system. Therefore, operating at the EP $\Omega^{(1)}_\textrm{EP}$, the coupling strength between oscillators can be estimated from the peaks of optical spectrum.



\section{Conclusion}

We investigated the optical spectrum of a trimer chain driven by a quadratic photon. We have shown the emergence of multiple exceptional points corresponding to the coalescence of eigenvalues and eigenvectors of the dynamical matrix. In a particular exceptional point $\Omega=\Delta$, the resulted optical spectra are expressed as a product of a conventional Lorenztian spectrum with a dispersive term of $\omega^4$, which is absent in the case of a single oscillator. 
Unlike the case of a single oscillator, the presence of coupling gives rise to the emergence of two peaks, instead of a single peak. In particular, those two peaks are located at the natural frequency of the trimer in a closed system of $\omega=\pm\sqrt{2}J$ proportional to the coupling strength $J$, regardless of the spectra of the eigenvalues. Furthermore, we have found that these peak positions persist for $\Omega>\Delta$. Therefore, the coupling between the oscillator can be estimated from the splitting of the two peaks. 
\begin{acknowledgments}
We thank Mahameru BRIN for their HPC facilities.  A. A. is supported by research assistantship from BRIN Directorate for Talent Management.  We thank Dr. Charles Downing for a fruitful discussion.
\end{acknowledgments}


%

\end{document}